# Any Regulation of Risk Increases Risk

**Philip Z. Maymin and Zakhar G. Maymin**

Dr. Philip Z. Maymin is Assistant Professor of Finance and Risk Engineering at NYU-Polytechnic Institute. He is also the founding managing editor of *Algorithmic Finance*.

He holds a Ph.D. in Finance from the University of Chicago, a Master's in Applied Mathematics from Harvard University, and a Bachelor's in Computer Science from Harvard University. He also holds a J.D. and is an attorney-at-law admitted to practice in California.

He has been a portfolio manager at Long-Term Capital Management, Ellington Management Group, and his own hedge fund, Maymin Capital Management. He has also been a policy scholar, a Congressional candidate, and an award-winning journalist.

His textbook Financial Hacking is forthcoming by World Scientific.

Dr. Zakhar G. Maymin is CEO of Quantitative Investment Services.

He holds a Ph.D. in Mathematics from MIT and a Master's in Mathematics from Moscow University.

He has been a portfolio manager at Ellington Management Group and his own hedge fund, Maymin Capital Management. He has also been an Assistant Professor of Mathematics at Northeastern University, the Head of the North American New Products Group at Sakura Global Capital, and Senior Research Associate at Susquehanna Investment Group.

He was also the New York State chess co-champion and the Fairfield County (CT) open class chess champion. He is the author of Publicani and Ethics for Your Child.

Corresponding author:
Philip Z. Maymin
Assistant Professor of Finance and Risk Engineering
Department of Finance and Risk Engineering
NYU-Polytechnic Institute
Six MetroTech Center
Brooklyn, NY 11201
Phone: (718)260-3175
Fax: (718)260-3355
Email: philip@maymin.com

Zakhar G. Maymin
Email: zak@maymin.com

# Any Regulation of Risk Increases Risk

**Abstract:** We show that any objective risk measurement algorithm mandated by central banks for regulated financial entities will result in more risk being taken on by those financial entities than would otherwise be the case. Furthermore, the risks taken on by the regulated financial entities are far more systemically concentrated than they would have been otherwise, making the entire financial system more fragile. This result leaves three directions for the future of financial regulation: continue regulating by enforcing risk measurement algorithms at the cost of occasional severe crises, regulate more severely and subjectively by fully nationalizing all financial entities, or abolish all central banking regulations including deposit insurance to let risk be determined by the entities themselves and, ultimately, by their depositors through voluntary market transactions rather than by the taxpayers through enforced government participation.

1. **Introduction**

When depositors become anxious about the safety of their deposits in a particular bank, they rush to get their money out because the remaining assets are first-come, first-served, and if there is not enough for everyone, latecomers get nothing. Such bank runs force banks to liquidate whatever assets they had purchased with the deposits, often at a substantial loss. Instead of letting the free market determine how each bank invests its deposits, governments seek to avoid such runs altogether by offering deposit insurance: depositors are assured their money is safe no matter what. The existence of the deposit insurance makes people indifferent to the safety of their banks and so they don't bother running to get their money out even if their bank is on the brink of insolvency. Conversely, the lack of deposit insurance makes people concerned about safety and imposes discipline on the banks. Peria and Schmukler (2001) empirically document the existence of such market discipline on banks in the complete absence of deposit insurance, or when deposit insurance is not credible, in Argentina, Chile, Mexico during the 1980s and 1990s.

Ioannidou and de Dreu (2006) show that deposit insurance indeed causes a significant reduction in market discipline in Bolivia from 1998 to 2003, and that the effect of market discipline completely vanishes when insurance covers 100 percent of the possible loss.

The trouble is that deposit insurance removes the responsibility from the depositors to determine a safe location for their assets. Any insured bank becomes just as good as any other. The Federal Reserve Board and the Statistics of Income Division of the Internal Revenue Service sponsor a triennial Survey of Consumer Finances to provide, among other things, detailed information about the respondents' use of financial services. Bucks, Kennickell, Mach, and Moore (2009) report that for the four such surveys conducted in 1998, 2001, 2004, and 2007, the number one response consumers cited as the most important reason for choosing their primary financial institution was the location of the bank's offices, with more than 40 percent of respondents indicating geographical convenience as their most important reason. The second and third most popular reasons were low fees and the ability to obtain many services at one place, at about 15 percent response each. Safety and the absence of risk were listed as next to last, at only 2 percent on average. In short, with deposit insurance, consumers are indifferent as to the particular risk each bank runs.

What does such consumer indifference to bank risk mean for the banks? What would you do if you could start a bank with deposits insured by the government? You might invest prudently and grow your business responsibly. Or you might buy lottery tickets: if you win, you keep virtually all of the profits, and if you lose, you have no personal liability anyway. And as we saw from the Survey of Consumer Finances, you are likely to be able to attract depositors simply by offering convenient locations and low fees because customers don't care what kind of risk you take: even if you lose, they will get their money back from the government.

Hendrickson and Nichols (2001) compared Canadian and American bank data in a historical study to conclude that deposit insurance does indeed increase risk taking. Calem and Rob (1999) show that even a deposit insurance surcharge does not deter banks near insolvency from increasing risk.

But governments can't just blindly guarantee the deposits of any financial institution. They can't allow reckless risk taking. So what can they do? They must institute restrictions on banks in order for them to qualify for the deposit insurance and its ancillary benefits such as overnight lending and borrowing of excess funds and other clearing operations.

Chief among the restrictions that governments and central banks place on individual banks is the amount of risk capital that banks must allocate to support a position. For example, if you as a bank want to buy $100 worth of IBM stock, how much risk capital do you need to reserve such that you are able to withstand extreme losses without being forced to liquidate under duress?

The most commonly used approach in evaluating the reserve requirement is Value-at-Risk (VaR). Though the VaR is technically the worst-case loss of a truncated historical distribution, it is at heart merely a multiple of the standard deviation of the portfolio value, for two reasons. First, regulations often allow banks to use parametric estimation methods based on continuous distributions, such as the Gaussian, so long as they apply an additional multiple on their resulting number to compensate for fat tails. Second, suppose that the VaR is computed non-parametrically using actual distribution history, and that the returns are generated from a non-Gaussian distribution with an excess kurtosis. Even in that situation, the VaR is still usually some relatively stable multiple of the standard deviation.

Thus, governments allow a bank to hold a portfolio so long as the bank sets aside risk capital equal to some constant $c$ times the historical standard deviation of that particular portfolio. This algorithm has undergone some evolution from the original "Basel I" agreement of 1988 through the most recent changes ("Basel II") and the new changes now being phased in ("Basel 2.5") and expected to take effect in 2013 ("Based III"), with the most recent changes coming in the wake of the global financial crisis. Broadly, the evolution in regulations has gone from a constant depending only on asset class, to depending on the current VaR, to (most recently) depending on the sum of the current VaR and the worst-case VaR during a historical stress period for the asset under consideration, such as 2008-2009.

The standard requirement called for risk capital of three times the 10-day 99% VaR. The 10-day standard deviation is about $\sqrt{10} = 3.16$ times the daily standard deviation. The 99% VaR is about 2.33 standard deviations, from the inverse cumulative distribution function of the standard normal. So the standard requirement in effect called for a market risk reserve requirement for a portfolio of $c = 3.16 * 2.33 * 3 = 22$ times the portfolio's historical daily standard deviation. For portfolios that historically moved about one percent per day on average, the risk requirement would have been about twenty-two percent.

In any event, the regulation has always been some function over past prices and returns. The form of the function may change but unless the regulator personally inspects and approves each possible trade or portfolio, thus effectively nationalizing all financial services and bringing them under purely political control, the only way to regulate is to provide a list of clear and objective rules.

The problem is that any such rules to reduce risk will result in more risk. The crux of this paper consists of the following argument. Any objective risk regulation rules discriminate among investment opportunities in the sense that some investments become more attractive than others based on the formal regulatory algorithm. Any such discrimination leads to a distortion of investment opportunities because banks will tend to switch into the more favored investments, and, finally, any such distortion leads to increased individual and systemic risk.

Duchin and Sosyura (2011) document that bailed-out banks subsequently increase risk, and they do so within asset classes, so that the increased riskiness is less apparent. Further, they note that the bailed-out banks subsequently appear safer according to capitalization requirements, but in fact are much riskier. Thus, they conclude that the response by banks to capital requirements may hinder the efficacy of risk regulation. Going further, we show here by a more general argument that in fact *any* risk regulation will ultimately result in more systemic risk.

Much literature has focused examination on big banks and on comparing various possible regulatory changes. Saunders and Walter (2012) note that one implication of risk socialization is the likelihood of certain institutions becoming too big or too interconnected to be allowed to fail. Gatzert and Schmeiser (2011) find that the benefits commonly attributed to risk diversification within financial conglomerates is often overstated. Cao and Illing (2010) analyze optimal regulatory responses in the presence of systemic liquidity shocks. By contrast, we do not focus only on the biggest banks or on marginally different regulatory proposals, but rather on the more fundamental questions of whether or not regulation itself can succeed in reducing systemic risk, regardless of the size of the banks.

Hermsen (2010) points out that Basel regulations do not specify a method of calculating VaR, and hence banks will choose that VaR algorithm that allows them to establish riskier positions. Our approach here is similar in spirit in that we agree that banks will change their portfolio allocation algorithm in response to the regulatory rules. However, Hermsen's argument can be entirely addressed by a new Basel accord stipulating either more restrictions on the choice of VaR model or explicitly specifying how VaR must be calculated. Our finding, on the contrary, continues to hold regardless of whatever other regulations are passed, even if the VaR model is explicitly specified.

Kaplanski and Levy (2007) culminate a long line of literature that assumes expected utility maximizing banks in a mean-variance framework to examine the effect of VaR regulation. They find that there is an optimal VaR-based regulation, although current Basel levels exceed that optimal amount. Yet we will show that *any* regulation will result in more risk. Why the discrepancy? Kaplanski and Levy (2007) assume that regulated banks continue to act in a mean-variance world; in other words, regulations result in a change in allocations but no change in the allocation algorithm. Here, however, we argue that there is a substantial behavioral change between no regulation and some regulation that results in banks changing the way they determine their portfolio.

The differences between our conclusions and those of Kaplanski and Levy (2007) can be illustrated in our main example on VaR-based regulation. They assume that the true future variance is known and examine the changes in the mean-variance frontier induced by regulations. Here we argue that banks will change their portfolio based on the random value of the estimate of variance, that the banks will be systematically biased towards those securities whose variance erroneously and randomly appears low, that the future variance of those securities should theoretically be higher, and that, as predicted, the future variance is indeed higher in empirical tests.

## 2. Theory and Calculation

For the simplest case, imagine there are *m* identical securities, each of which has returns that are independently normally distributed with zero mean and true standard deviation σ, and we have a history of *n* periods for each of them. Think of *m* as being a few thousand securities and of *n* as being about sixty monthly returns, or five years of data.

What is the distribution of the sample standard deviations $s_i$ of each of the securities? Simply by chance, about half of the securities will have sample standard deviations above σ, and half below. More specifically, the sample standard deviations follow a $\chi^2$ distribution, under which the probability of a sample standard deviation being below σ is greater than one-half and decreases to its limiting value of one-half as the number of observations increases. For *m* = 1000 and *n* = 60, we should expect about ten securities to exhibit a sample standard deviation less than 80 percent of its true standard deviation σ, regardless of the particular value of σ. Let us prove a specific theorem.

### 2.1. Theorem: The Conditional Expected Value of the Sample Standard Deviation

Suppose that $y_i, i = 1,2,\ldots,n$, are independent and identically distributed $\mathcal{N}(\mu, \sigma^2)$ normal returns and the sample standard variances $s_n^2$ are defined as usual by:

$$s_n^2 = \frac{1}{n-1} \sum_{i=1}^{n} (y_i - \bar{y})^2$$

where:

$$\bar{y} = \frac{1}{n} \sum_{i=1}^{n} y_i$$

Then we can calculate the conditional expected value of the sample standard deviation as a percentage of the true standard deviation for any level $\alpha$ between 0 and 1 as follows:

$$\frac{E(s_n \mid s_n \leq s_{n,\alpha})}{\sigma} = K_n \frac{P(\chi_n^2 \leq \chi_{n-1,\alpha}^2)}{\alpha} \quad (1)$$

where:

$s_{n,\alpha}$ is the number such that $P(s_n \leq s_{n,\alpha}) = \alpha$, \quad (2)

$\chi_{n-1,\alpha}^2$ is the number such that $P(\chi_{n-1}^2 \leq \chi_{n-1,\alpha}^2) = \alpha$, \quad (3)

$\chi_n^2$ is the chi-square random variable with $n$ degrees of freedom,

$$K_n = \frac{\Gamma\left(\frac{n}{2}\right)}{\Gamma\left(\frac{n-1}{2}\right)\left(\frac{n-1}{2}\right)^{\frac{1}{2}}}, \text{ and} \quad (4)$$

$\Gamma(x)$ is the gamma function.

Similar results hold for higher tails by substituting "$\geq$" everywhere for "$\leq$".

Furthermore, a constant $K_n$ converges quickly to 1:

$$\lim_{n \to \infty} K_n = 1 \quad (5)$$

*2.2. Proof of Theorem 2.1*

Let's find the p.d.f. of $s_n$. First, because the ratio of the sample variance to the true variance is distributed as $\chi^2$,

$$(n-1)\frac{s_n^2}{\sigma^2} \sim \chi_{n-1}^2$$

its c.d.f. is:

$$F_{s_n}(x) = P(s_n \leq x) = P(s_n^2 \leq x^2) = P\left(\chi_{n-1}^2 \leq \frac{n-1}{\sigma^2}x^2\right) = F_{\chi_{n-1}^2}\left(\frac{n-1}{\sigma^2}x^2\right) \quad (6)$$

where $F_{\chi_{n-1}^2}(x)$ is the c.d.f. of $\chi_{n-1}^2$.

Therefore the p.d.f. of $s_n$ is:

$$f_{s_n}(x) = \frac{d}{dx}F_{s_n}(x) = \frac{2(n-1)x}{\sigma^2} f_{\chi_{n-1}^2}\left(\frac{n-1}{\sigma^2}x^2\right)$$

where $f_{\chi_{n-1}^2}(x)$ is the p.d.f. of $\chi_{n-1}^2$:

$$f_{\chi^2_{n-1}}(x) = \frac{1}{2^{\frac{n-1}{2}} \Gamma\left(\frac{n-1}{2}\right)} x^{\frac{n-1}{2}-1} e^{-x/2}, \qquad x \geq 0$$

We can rewrite $f_{s_n}(x)$ as:

$$f_{s_n}(x) = 2 \left(\frac{n-1}{2}\right)^{\frac{n-1}{2}} \frac{1}{\Gamma\left(\frac{n-1}{2}\right)\sigma^{n-1}} x^{n-2} e^{-\frac{(n-1)x^2}{2\sigma^2}} \tag{7}$$

Now, let's find the expected value of the sample standard deviation, conditional on it being in a low percentile.

$$E(s_n \mid s_n \leq s_{n,\alpha}) = \frac{1}{\alpha} \int_0^{s_{n,\alpha}} 2 \left(\frac{n-1}{2}\right)^{\frac{n-1}{2}} \frac{1}{\Gamma\left(\frac{n-1}{2}\right)\sigma^{n-1}} x^{n-1} e^{-\frac{(n-1)x^2}{2\sigma^2}} dx$$

Using the substitution $(n-1)x^2 = ny^2$, we get:

$$E(s_n \mid s_n \leq s_{n,\alpha}) = K_n \frac{\sigma}{\alpha} \int_0^{\sqrt{\frac{n-1}{n}} s_{n,\alpha}} 2 \left(\frac{n}{2}\right)^{\frac{n}{2}} \frac{1}{\Gamma\left(\frac{n}{2}\right)\sigma^n} y^{n-1} e^{-\frac{ny^2}{2\sigma^2}} dy$$

From equation (7) we see that the last expression has the p.d.f. of $s_{n+1}$ under the integral sign. So we can rewrite it as

$$E(s_n \mid s_n \leq s_{n,\alpha}) = K_n \frac{\sigma}{\alpha} P\left(s_{n+1} \leq \sqrt{\frac{n-1}{n}} s_{n,\alpha}\right) \tag{8}$$

To calculate the probability on the right hand side, we use equation (1) to write:

$$\alpha = P(s_n \leq s_{n,\alpha}) = P(s_n^2 \leq s_{n,\alpha}^2) = P\left(\frac{\sigma^2}{n-1} \chi^2_{n-1} \leq s_{n,\alpha}^2\right) = P\left(\chi^2_{n-1} \leq \frac{n-1}{\sigma^2} s_{n,\alpha}^2\right)$$

thus showing that:

$$\frac{n-1}{\sigma^2} s_{n,\alpha}^2 = \chi^2_{n-1,\alpha}$$

Therefore we can rewrite equation (8) as:

$$E(s_n|s_n \leq s_{n,\alpha}) = K_n \frac{\sigma}{\alpha} P\left(s_{n+1}^2 \leq \frac{n-1}{n} s_{n,\alpha}^2\right) = K_n \frac{\sigma}{\alpha} P\left(\frac{\sigma^2}{n} \chi_n^2 \leq \frac{n-1}{n} s_{n,\alpha}^2\right)$$

$$= K_n \frac{\sigma}{\alpha} P\left(\chi_n^2 \leq \frac{n-1}{\sigma^2} s_{n,\alpha}^2\right) = K_n \frac{\sigma}{\alpha} P(\chi_n^2 \leq \chi_{n-1,\alpha}^2)$$

thus proving the theorem in equation (1).

To prove that $K_n \to 1$, we note that according to Stirling's formula:

$$\lim_{z \to \infty} \frac{\Gamma(z)}{\sqrt{2\pi} z^z e^{-z} z^{-\frac{1}{2}}} = 1$$

To take advantage of this formula as follows, we rewrite equation (4):

$$K_n = \frac{\Gamma\left(\frac{n}{2}\right)}{\Gamma\left(\frac{n-1}{2}\right) \left(\frac{n-1}{2}\right)^{\frac{1}{2}}} = A \cdot B \cdot C$$

where

$$A = \frac{\Gamma\left(\frac{n}{2}\right)}{\sqrt{2\pi} \left(\frac{n}{2}\right)^{\frac{n}{2}} e^{-\frac{n}{2}} \left(\frac{n}{2}\right)^{\frac{1}{2}}} \quad B = \frac{\sqrt{2\pi} \left(\frac{n-1}{2}\right)^{\frac{n-1}{2}} e^{-\frac{n-1}{2}} \left(\frac{n-1}{2}\right)^{\frac{1}{2}}}{\Gamma\left(\frac{n-1}{2}\right)} \quad C = \frac{e^{-\frac{n}{2}} \left(\frac{n}{2}\right)^{\frac{1}{2}}}{e^{-\frac{n-1}{2}} \left(\frac{n-1}{2}\right)^{\frac{1}{2}}} \cdot \left(\frac{n}{n-1}\right)^{\frac{n}{2}}$$

and by Stirling's formula $A \to 1$, $B \to 1$, and

$$C = e^{-\frac{1}{2}} \left(\frac{n}{n-1}\right)^{\frac{1}{2}} \cdot \left(\frac{n}{n-1}\right)^{\frac{n-1}{2}} \cdot \left(\frac{n}{n-1}\right)^{\frac{1}{2}} \to 1$$

because it is well known that:

$$\lim_{n \to \infty} \left(\frac{n+1}{n}\right)^n = e$$

Therefore $K_n \to 1$ and this completes the proof. Indeed, the convergence is quick, as even for $n = 60$, $K_n = 0.996$.

*2.3. Response of the Banks*

Suppose that the government-mandated risk capital requirement for each security is some constant *c* times that security's sample standard deviation. What would be the natural response of the banks?

Much like the lottery ticket bank illustrated above, banks would tend towards buying portfolios that are riskier than they appear. A particular security that had a sample standard deviation of eighty percent of the true standard deviation would let banks spend twenty percent less risk capital that they ought to while maintaining a full exposure to the true risk. And there would on average be approximately ten such securities at any given point in time.

And that is the basic idea. Purely by random chance, a few securities will appear to have much lower risk than they truly do. Banks will gravitate towards establishing positions in these securities because they are able to use less risk capital on them than on arbitrary average positions, and banks do not face significant market discipline for establishing too much risk because the government guarantees deposits. Therefore, instead of different banks simply holding different well-capitalized risky positions, all banks will tend to hold combinations of those few and rare securities that falsely appear to have decreased risk. If banks and their customers had to bear this risk, they would avoid holding securities with such low risk reserves; they do it only because it is not their risk or their customers', the algorithm set by the regulator allows it, and they are able to earn higher returns on capital through this use of excess leverage.

While other regulations may aim to preclude each individual bank from concentrating its holdings into a few securities, there is no regulation short of nationalization that could preclude all banks from investing in the same few assets. So the effect of an algorithmic approach to determining risk capital is that all banks will tend to establish the maximum position they possibly can into the very few securities that randomly exhibited lower risk, and thus lower required risk capital, than they should have.

Thus, when any one of these particular securities later experience a typical downwards movement, it will appear to be a significantly aberrant move from the perspective of the required risk capital and the observed (low) historical sample standard deviation, requiring the banks to liquidate those and other positions quickly to raise enough cash to replenish their reserves. And so a relatively modest move in a few key securities could suddenly result in the collapse of the entire financial system.

One might argue that because each asset has an unobserved true volatility and an observed empirical volatility and nobody knows the relation between the two, it is not possible to say if the observed empirical volatility is "too low," and hence the kind of position-picking by banks described here would not likely take place. In other words, how do we know when we observe a low empirical volatility that it is not actually an accurate, or perhaps even overstated, reflection of the true volatility? The answer is that the likelihood of this happening is scarce: most likely, the lowest observed volatility assets exhibit such low numbers because of noise. We quantified this claim in Theorem 2.1 and will examine it empirically in the next section.

*2.4. Empirical Evidence*

Is there any empirical evidence for this effect? We use the CRSP database of all stocks listed on the NYSE, AMEX, and Nasdaq (after 1972) and, for each date from January 1932 through December 2003, calculate the standard deviation of the sixty monthly returns for the five years prior, and the standard deviation of the sixty monthly returns for the five year after that date. For securities whose history ends less than five years after the date, we use the standard deviation of however many monthly returns are available. For each stock we calculate the ratio of the new standard deviation to the old standard deviation. Then on each date we sort the stocks based on past standard deviations and plot for different quantile groups the associated ratio of the new standard deviation to the old standard deviation.

Exhibit 1 plots the time series of ratios of new standard deviations to old standard deviations for five such quantile groups: for the lowest 1 percent of past standard deviations, for the range from 1 percent to 10 percent, for the range from 10 percent to 90 percent, for the range from 90 percent to 99 percent, and for the range from 99 percent to 100 percent. Note that each group always lies above the next ones. The overall averages for the five groups are, respectively, 1.85, 1.17, 1.00, 0.81, and 0.56. In other words, stocks with the lowest one percent of past five-year standard deviations on average experience an 85 percent higher standard deviation in the subsequent five years. Thus, empirically as well as theoretically, stocks that look too good to be true usually are.

As banks begin to prefer and invest in the regulatorily favored assets, the prices of those assets should increase, at least temporarily through their market impact. Indeed, the low-volatility puzzle of Ang, Hodrick, Xing, and Zhang (2009) in which assets recently exhibiting lower volatility tend to have higher future returns.

## 3. Results and Discussion

One possible response by regulators to this observation is to require the use of more data, either by looking further back in time or by requiring the use of more frequent observations.

The two problems with looking further back are with existence and consistency. Many securities simply don't have that much history so the longer regulators require the historical lookback to be, the greater the penalty for younger securities. Furthermore, the character of a particular security, especially a stock, may have changed substantially from the kind of company it was many years ago, either because of a change in business focus, or because of a change in the risk factor loadings and characteristics of the stock due to growth in market capitalization or revenues or a change in value.

So the alternative remains to require more frequent observations over the same time period, for example requiring the use of daily returns rather than monthly returns. Over five years, that means that $n$, the number of periods, increases from 60 to 252 * 5 = 1,260, assuming 252 business days per year on average.

As Exhibit 2 demonstrates, continuing our assumption of $m = 1000$ independent securities but increasing the number of periods $n$ does indeed decrease the expected value of the sample standard deviation in the bottom one percent. Theorem (2.1) derives a convenient formula for calculating such an expectation. For $n = 1260$, the conditional expected value of the 0.1 percent tail is 0.93 times the true standard deviation, meaning even the biggest deviations in sample standard deviation are on average within seven percentage points of the true risk.

Does this mean that increasing the number of observation periods solves all of the problems of a concentration of risk by banks into a handful of seemingly less risky securities?

For two reasons, no. First, even a small difference between the required regulatory risk capital and the true risk of the position may entice banks to prefer such securities over other securities. But secondly, the above analysis misses an important aspect of real-world returns: they are not normal.

One obvious deviation from normality in security returns is fat tails. The normal distribution has an index of kurtosis, the standardized fourth central moment, of exactly three. But observed kurtosis often exceeds ten times that amount. For example, the kurtosis of the daily S&P 500 index returns since 1950 is 25.8.

We can simulate fat-tailed returns with the following simple approximation algorithm: draw from a standard normal distribution but replace all draws within some small constant $\varepsilon$ of zero with a large constant jump $h$ of the same sign as the original draw. For example, if $\varepsilon = 0.01$ and $h = 10$, then a random draw of -0.005 would be replaced by -10.

Simulating 1,000,000 such random numbers gives a distribution with a near-zero mean (0.002 compared with 0.000 for the standard normal), a very slightly elevated standard deviation (1.34 compared to 1.00 for the standard normal), but a very high kurtosis (25.6 compared to 3 for the standard normal). In other words, if we standardize the results by dividing by its 1.34 standard deviation, we are able to use this algorithm to simulate fat-tailed returns.

Using this distribution, we calculate 1,000 different sample standard deviations for 1,260 observation periods, the equivalent of five years of daily returns. Exhibit 3 shows the histogram of standardized simulated standard deviations. Note that a few securities are near the 0.80 ratio again, meaning that adding kurtosis to the distribution to better match empirical returns offsets the additional accuracy resulting from using more observation periods.

Thus, for reasonable values of the number of securities $m$ and the number of observation periods $n$, there will still emerge a handful of stocks that appear less risky than they actually are, and which will attract banks to hold them for the same reasons outlined above.

Could the new requirements of Basel II to add an additional stress term alleviate this problem? We can answer this question through simulation as well. Simulate 1,260 returns for each of 1,000 securities using the high kurtosis algorithm described above. For each sequence of returns, compute both the overall five-year standard deviation and the highest rolling yearly standard deviation, and report the sum. Basel II effectively requires a multiple of this sum as risk capital. Exhibit 4 displays the histogram of these standardized values as well. Again, there emerge several securities purely by randomness that appear to have much lower overall risk than the average security. So the addition by Basel II of "stress" risk does not alleviate the problem.

The risk incentive problem depends implicitly on the limited liability of shareholders and their call option-like payoff structure. Thus, even risk averse shareholders with diversified investment portfolios will force managers to increase risk whenever depositors are indifferent.

Depositors are indifferent if their deposits are insured, such as with the FDIC. Could slight changes to the nature of the FDIC alleviate the problem? Limits on the amounts insured only matter when they become so low that the transactions costs of opening multiple accounts becomes too expensive; thus any reasonable limits on the insurance are in essence the same as full insurance. On the other hand, if the regulatorily required fees for deposit insurance from each bank only acted to reduce the losses of the insured for that bank, then there would be virtually no insurance and depositors would be wary of where they placed their money, and the possibility of bank runs would immediately return.

What about asset classes for which the past does not represent their full risk? Banks may prefer assets whose risks are not captured by historical datasets, such as tail risks and out-of-the-money options where losses are hidden, infrequent, and large. This concern is real and true and only serves to exacerbate the problems we describe. The problem is large even when ignoring hidden risks, but our approach is conservative and the real risk is larger still.

Expected returns are also difficult to estimate. Could the two errors offset each other? For example, if assets with lower estimates of volatility routinely also showed lower expected returns, then banks would be less prone to overinvest in them than otherwise. However, there are five reasons why this is not likely. First, as Merton (1980) shows, expected return are even harder to estimate precisely than standard deviations. Second, expected returns tend to be less persistent than volatilities. Third, expected excess returns tend not to be statistically significantly different from zero in practice. Fourth, even if some assets that appear less risky than they truly are also exhibit expected returns that are lower than they truly are, the expected returns would need to be even more understated than the volatility in order for the investment to be less attractive. Fifth, there would still be other assets that appear less risky than they truly are whose expected returns are not also expected to be lower. Thus, the estimation errors of expected returns will not offset the estimation errors of risk.

What about other forms of risk regulation, for instance, if risk regulation additionally punished non-diversification, to attempt to discourage overinvestment in the favored assets? Then there will still exist some measure according to which some investments or sets of investments will appear more attractive to the regulators due solely to randomness. This is true for any value function proposed by the regulators. Different value functions produce different risks, but the main argument holds: any such measure will introduce additional risk because it gives an artificial incentive for the banks to invest into some regulatorily favored securities that may look attractive only because of inherent randomness. In other words, different regulations may cause different assets to be favored, but there will always be some that differ from their true risk. On the other hand, if there are no regulations, the risk of randomness gets more diversified across assets and more dispersed across individual banks, thus lowering systemic risk.

## 4. Conclusion

The risk capital that banks allocate to their positions must be set by a regulator if deposits are insured. If, as a first option, the regulator determines the appropriate risk capital on a case-by-case basis, then banks are effectively nationalized and run by the regulator. This has often not seemed like a palatable choice, and so regulators have attempted, as a second option, to craft seemingly objective risk measurement rules that banks are required to follow.

The most common kind of risk measurement rules have been variations of measures of standard deviation, and it turns out that all such rules, for any reasonable history lengths, encourage banks to invest in a few securities that are riskier than they appear, thus increasing systemic risk.

The results are even more general, though. Even if regulators required a risk capital reserve of 100 percent of market value for every security, the risk reserve still would not match the true risk of each security, because the price of a security is not necessarily its risk, and so banks would still tend to invest in the same few riskier assets.

Thus, we can conclude that the effect of *any* objective rules for regulation will result both in more risk being taken by each individual bank, and by the risks taken by different banks to be more correlated with each other, resulting in a far more fragile financial system than would be the case otherwise.

The only third option is to not regulate at all, and to not insure deposits. This would leave each bank, and its customers and depositors, with the ultimate responsibility of determining the appropriate risk capital. This option deserves greater consideration in light of the results of this paper. In the meantime, portfolio managers, risk managers, policymakers, regulators, and taxpayers ought to be aware of this previously unknown source of risk.

**Acknowledgments**

The authors are grateful for comments and suggestions by the anonymous referee, Darrell Duffie, David R. Henderson, participants at the Stanford University Workshop on Capitalism's Crises, the Society of Actuaries conference, and a research presentation at Oliver Wyman.

# Figure 1: Time Series of Volatility Ratios

The ratios of future five-year monthly standard deviation to past five-year monthly standard deviation, arranged into five quantile groups by past standard deviations, is plotted on a log scale. Stocks with recent low standard deviation tended to have higher standard deviations going forward, and vice versa.

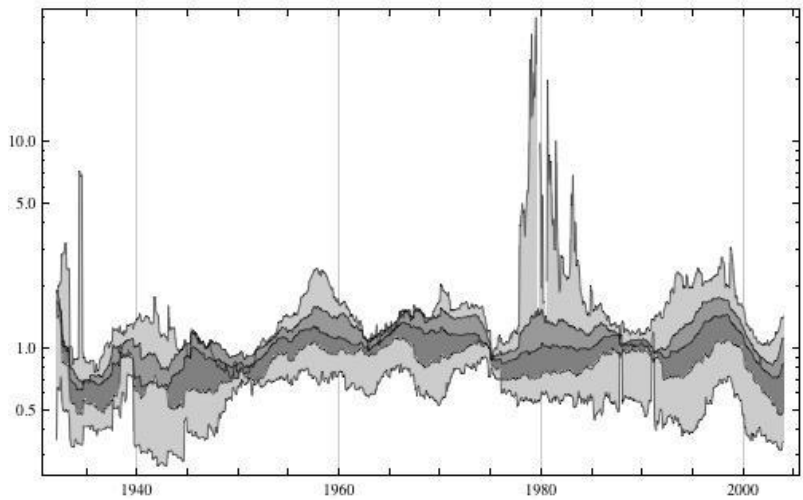

## Figure 2: Expected Tail Sample Standard Deviations

The formula for deriving these conditional expected values of the ratio of the sample standard deviation to the true standard deviation is derived in Theorem (2.1) and plotted here for a number of periods ranging from 30 to 1200 and for tail probabilities of 1 percent (top line) and 0.1 percent (bottom line). For example, for 60 time periods, the expected values of the standard deviation are 0.76 and 0.71 times the true standard deviation for the two probabilities respectively.

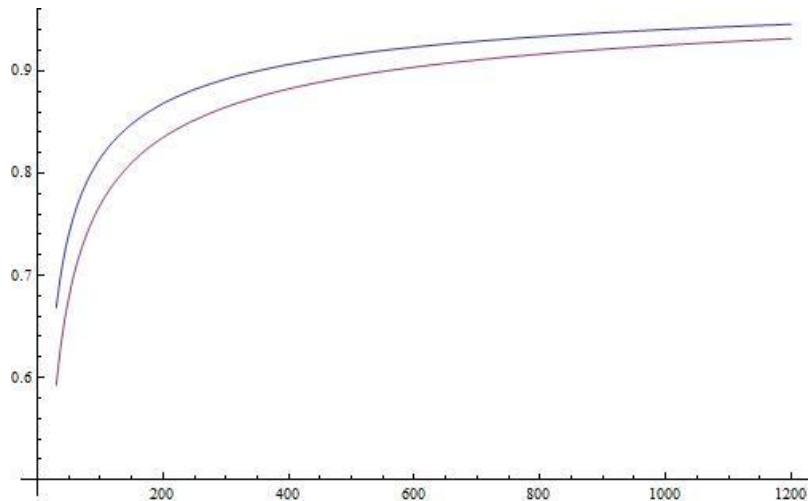

# Figure 3: Simulated Standard Deviations or Basel I Risk

There exist several securities in the left part of the histogram near the ratio of 0.8, meaning that there will still exist by pure chance some small number of securities that appear to have less risk than they truly do.

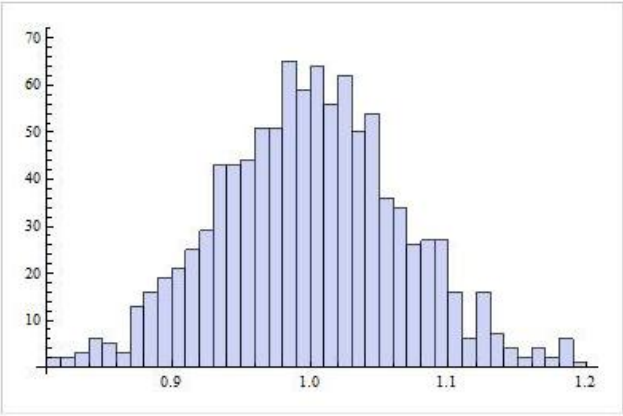

## Figure 4: Simulated Basel II Risk

There exist several securities in the left part of the histogram near the ratio of 0.8, meaning that even with the addition of the maximum of a rolling standard deviation to the usual standard deviation of returns, there will still exist by pure chance some small number of securities that appear to have less risk than they truly do.

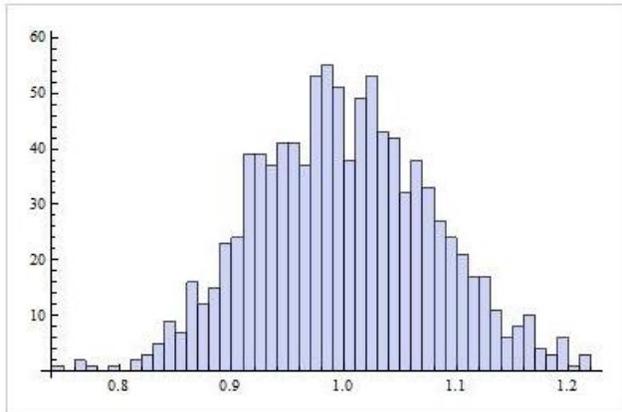